# Beam-driven ECH waves: A parametric study

Xu Zhang[1], Vassilis Angelopoulos[1], Anton V. Artemyev[1], Xiao-Jia Zhang[1]

[1]Department of Earth, Planetary, and Space Sciences, and Institute of Geophysics and Planetary Physics, University of California, Los Angeles, California, USA

Abstract: Electron cyclotron harmonic (ECH) waves play a significant role in driving the diffuse aurora, which constitutes more than 75% of the particle energy input into the ionosphere. ECH waves in magnetospheric plasmas have long been thought to be excited predominantly by the loss cone anisotropy (velocity-space gradients) that arises naturally in a planetary dipole field. Recent THEMIS observations, however, indicate that an electron beam can also excite such waves in Earth's magnetotail. The ambient and beam plasma conditions under which electron beam excitation can take place are unknown. Knowledge of such conditions would allow us to further explore the relative contribution of this excitation mechanism to ECH wave scattering of magnetospheric electrons at Earth and the outer planets. Using the hot plasma dispersion relation, we address the nature of beam-driven ECH waves and conduct a comprehensive parametric survey of this instability. We find that growth is provided by beam electron cyclotron resonances of both first and higher orders. We also find that these waves are unstable under a wide range of plasma conditions. The growth rate increases with beam density, beam velocity, and hot electron temperature; it decreases with increasing beam temperature and beam temperature anisotropy ($T_\perp/T_\parallel$), hot electron density, and cold electron density and temperature. Such conditions abound in Earth's magnetotail, where magnetospheric electrons heated by earthward convection and magnetic reconnection coexist with colder ionospheric electrons.

# 1. Introduction

An electron cyclotron harmonic (ECH) wave is an electrostatic emission in the $nf_{ce}$ (electron cyclotron frequency; $n = 1,2,3 ...$) to $(n + 1)f_{ce}$ frequency range with the strongest wave power in its first harmonic band, $n = 1$ [Fredricks and Scarf, 1973; Shaw and Gurnett, 1975; Roeder and Koons, 1989; Meredith et al., 2009; Zhang and Angelopoulos, 2014]. First observed by the OGO 5 satellite, such electrostatic waves have wave power centered around $(n + 1/2)f_{ce}$ [Kennel et al., 1970]. They have since been observed over a large range of radial distances in both Earth's inner magnetosphere [Meredith et al., 2009; Ni et al., 2011a, 2017] and its magnetotail [Liang et al., 2011; Zhang et al., 2014]. The most intense ECH waves in Earth's magnetosphere, which are often within a few degrees in latitude relative to the magnetic equator, have been frequently observed between 2100 and 0600 magnetic local time, i.e. in the night-side magnetosphere [Roeder and Koons, 1989; Meredith et al., 2009; Ni et al., 2017]. They have also been observed in the magnetospheres of other planets, such as Jupiter [Kurth et al., 1980; Menietti et al., 2012] and Saturn [Gurnett et al., 2005; Tao et al., 2010] ], and in active space experiments [Mourenas et al., 1989]. In Saturn's magnetosphere, ECH waves tend to intensify and have more harmonic bands during injection events [Menietti et al., 2008; Tao et al., 2010]. Because both ECH waves and whistler-mode waves can resonate with electrons over the broad energy range from hundreds of eV to tens of keV, the relative importance of these waves in driving diffuse aurora had been controversial for decades [Kennel et al., 1970; Lyons, 1974; Belmont et al., 1983; Horne and Thorne, 2000; Horne et al., 2003; Meredith et al., 2000, 2009; Thorne et al., 2010]. Recently, however, it was recognized that ECH waves play an important role in scattering plasma electrons into the loss cone and in driving diffuse aurora in the outer magnetosphere, beyond eight Earth radii in the magnetotail [Ni et al., 2011b, 2012, 2016, Zhang et al., 2013, 2015].

Previous theoretical work demonstrated that ECH waves can be excited by the loss-cone distribution with a positive phase space density slope perpendicular to the ambient magnetic field ($\partial f / \partial v_\perp > 0$) [Young et al., 1973; Karpman et al., 1975; Ashour-Abdalla and Kennel, 1978; Ashour-Abdalla et al., 1979]. Assuming an unstable loss-cone distribution of a hot electron component in the presence of a cold electron component, Ashour-Abdalla and Kennel [1978] demonstrated that when the density and temperature ratios of cold to hot electrons are small enough, ECH waves are unstable. Electron cyclotron harmonic waves driven by loss-cone distributions usually propagate at very large (around 88°~ 89°) wave normal angles with respect to the background magnetic field and are heavily damped at smaller wave normal angles by Landau resonance [Horne, 1989; Mourenas and Beghin 1991; Horne and Thorne, 2000; Horne et al., 2003; Ni et al., 2011b, 2012; Liu et al., 2018]. Because measuring electron distribution functions around and within the loss cone (from a few degrees to less than one degree in pitch-angle space) is difficult, excitation of ECH waves by loss-cone instability has never been demonstrated directly using space measurements. Despite this lack of direct observational evidence of their excitation mechanism, there has been very little questioning or reconsideration of this mechanism since the theoretical work on loss-cone instability of many decades ago mentioned earlier. Whether there are other mechanisms for excitation of ECH waves apart from the loss cone instability had been unknown, until very recently.

The first observational evidence for ECH wave excitation in the magnetosphere by means other than the loss-cone instability was presented by Zhang et al. [2021]. Using data from the THEMIS mission [Angelopoulos, 2008] they found that ECH waves in the magnetotail can also be excited by low-energy electron beams. In the absence of loss-cone distributions, ECH waves driven by electron beams are unstable at moderately oblique wave normal angles (~$70^0$) [Menietti et al., 2002]. According to Zhang et al. [2021], such ECH waves, which have a strong wave electric field parallel to the magnetic field, can be quite frequently observed in the magnetotail behind sharp fronts of dipolarizing magnetic flux populated by hot plasma and embedded within fast flows (so-called dipolarization fronts; see, e.g., Runov et al. 2009). That these waves are correlated with parallel electron flux enhancement in the subthermal energy range suggests that they are likely driven unstable by low-energy electron beams. In Zhang et al. [2021], the authors also provided theoretical evidence for excitation of ECH waves by such beams in Earth's magnetotail. Using electron distribution functions with electron beams with realistic, plasma-sheet-like input parameters, the authors solved the dispersion relation and demonstrated that ECH waves can indeed by driven unstable by low-energy electron beams.

Even so, the generation mechanism of ECH waves driven by electron beams and the beam and ambient plasma conditions that favor excitation of beam-driven ECH waves still remain unclear. In this paper we extend the Zhang et al. [2021] instability analysis in two ways. First, we analyze the generation mechanism of beam-driven ECH waves by evaluating their dispersion relations and growth rates. We emphasize the relative importance of different resonances in a straightforward approach. Second, we perform a comprehensive parametric survey to quantify the dependence of these waves on various plasma parameters. We investigate the plasma conditions under which ECH waves might be driven unstable by electron beams. Our results could greatly improve our understanding of the growth and damping of beam-driven ECH waves under different plasma conditions and provide theoretical guidance for future work regarding beam-driven ECH waves in many space plasma environments.

Since generation of ECH waves by electron beams, has rarely been considered before we start with a rather detailed instability analysis in Section 2, where we also calculate the individual contributions of Landau resonance and cyclotron resonances to the growth rate. We also explore excitation of an ECH wave by electron beams at its second harmonic band. We follow, in Section 3, with a comprehensive parametric study. We evaluate the dependence of dispersion relations and the growth rate of beam-driven ECH waves on various plasma parameters, including density, drift velocity, temperature and temperature anisotropy of electron beams, density and temperature of the hot electron population, and density and temperature of the cold electron population. Section 4 summarizes our results.

## 2. Excitation of beam-driven ECH waves

In this section, we solve the hot plasma dispersion relations using the Waves in Homogeneous, Anisotropic Multi-component Plasmas code (WHAMP, see Ronnmark 1982) and analyze the growth rate of beam-driven ECH waves. The electron distribution function is represented as the sum of bi-Maxwellians:

$$f(v_\perp, v_\parallel) = \sum_i f_i = \sum_i \frac{n_i}{\pi^{2/3} \alpha_{\perp i}^2 \alpha_{\parallel i}} \exp\left(-\left(\frac{v_\parallel - v_{di}}{\alpha_{\parallel i}}\right)^2\right) \cdot \exp\left(-\frac{v_\perp^2}{\alpha_{\perp i}^2}\right) \qquad (1)$$

where the subscript $i$ is the $i$th component of the electron distribution function, $n_i$ is the electron number density, and $\alpha_{\parallel i}$ and $\alpha_{\perp i}$ are the thermal velocity of electrons in directions parallel and perpendicular to the magnetic field.

Table 1 lists the components of the electron distribution function: one hot component, one cold component, and two beam components. The temperature of the ion population is 5keV (typical temperature of magnetotail plasma) with no temperature anisotropy, and the ion population remains unchanged. The background magnetic field strength is 50nT and the electron plasma beta ($\beta_e = (\sum n_e k_B T_e)/(B^2/2\mu_0)$) is 0.08. The ratio of the electron plasma frequency to the electron cyclotron frequency $\omega_{pe}/\omega_{ce}$ is 5, and the ratio of the upper hybrid frequency to the electron cyclotron frequency $\omega_{uh}/\omega_{ce}$ is 5.1. These parameters are typical of the ECH wave generation region in Earth's plasma sheet. The two electron beam components drift in directions parallel and antiparallel, respectively, to the magnetic field. No electron component has loss-cone distributions. We modified systematically the density and temperature of the cold electron population starting from those in the electron distribution function used in Zhang et al. [2021]. These changes, which will be explained later, will benefit the parametric study in Section 3. The input parameters in the distribution function in Table 1 are used to calculate the results shown in Figure 1 and Figure 2.

Before embarking on the parametric study, we investigate how electrons resonate with ECH waves in velocity space in order to gain insight into the nature of the waves' excitation mechanism. The resonance condition for the nonrelativistic case is expressed as:

$$\omega - k_{\parallel} v_{\parallel} = n|\omega_{ce}| \qquad (2)$$

where $\omega$ is the wave frequency, $k_{\parallel}$ is the wave vector in the direction parallel to the magnetic field, and $n$ is the resonance harmonic number. Figure 1 shows the dispersion relation of beam-driven ECH waves in wave number space and the individual growth rate contributions from different resonance harmonic numbers $n$. As illustrated in Figures 1(a) and 1(b), electron cyclotron harmonic waves driven by electron beams at a temperature of 100eV are most unstable at wave normal angles of $55^0$ and at wave frequency of $1.1 f_{ce}$. To calculate growth rate contributions from Landau and cyclotron resonances, we modified the WHAMP program based on Equations (3) to (11). The dispersion relation for electrostatic waves can be simplified as below:

$$\mathcal{D} = k_{\perp}^2 \varepsilon_{xx} + k_{\perp} k_{\parallel}(\varepsilon_{xz} + \varepsilon_{zx}) + k_{\parallel}^2 \varepsilon_{zz} = 0 \qquad (3)$$

where $k_\parallel$ and $k_\perp$ are wave vectors parallel and perpendicular to the magnetic field, and $\varepsilon_{xx}$, $\varepsilon_{xz}$, $\varepsilon_{zx}$ and $\varepsilon_{zz}$, different elements in the dielectric tensor used when solving the hot plasma dispersion relation, are defined as:

$$\varepsilon_{xx} = 1 + \sum_i \frac{\omega_{pi}^2}{\omega^2} \sum_n \int \frac{v_\perp (n/\Lambda_i)^2 J_n^2(\Lambda_i)}{\omega - k_\parallel v_\parallel - n\Omega_{ci}} U d^3 v \tag{4}$$

$$\varepsilon_{xz} = \sum_i \frac{\omega_{pi}^2}{\omega^2} \sum_n \int \frac{v_\perp (n/\Lambda_i) J_n^2(\Lambda_i)}{\omega - k_\parallel v_\parallel - n\Omega_{ci}} W d^3 v \tag{5}$$

$$\varepsilon_{zx} = \sum_i \frac{\omega_{pi}^2}{\omega^2} \sum_n \int \frac{v_\parallel (n/\Lambda_i) J_n^2(\Lambda_i)}{\omega - k_\parallel v_\parallel - n\Omega_{ci}} U d^3 v \tag{6}$$

$$\varepsilon_{zz} = 1 + \sum_i \frac{\omega_{pi}^2}{\omega^2} \sum_n \int \frac{v_\parallel J_n^2(\Lambda_i)}{\omega - k_\parallel v_\parallel - n\Omega_{ci}} W d^3 v \tag{7}$$

with

$$\Lambda_i = \frac{k_\perp v_\perp}{\Omega_{ci}} \tag{8}$$

$$U = (\omega - k_\parallel v_\parallel) \frac{\partial f_{0i}}{\partial v_\perp} + v_\perp k_\parallel \frac{\partial f_{0i}}{\partial v_\parallel} \tag{9}$$

$$W = (\omega - n\Omega_{ci}) \frac{\partial f_{0i}}{\partial v_\parallel} + \frac{n\Omega_{ci} v_\parallel}{v_\perp} \frac{\partial f_{0i}}{\partial v_\perp} \tag{10}$$

The subscript $i$ is the $i$th component of the electron distribution function, $\omega_{pi}$ is the plasma frequency of the $i$th component, $n$ is the resonance harmonic number in Eq. (2), $\Omega_{ci}$ is the cyclotron frequency of the $i$th component, $J_n$ is the Bessel function of the first kind, and $f_{0i}$ is the phase space density of the $i$th component. Assuming weak growth or damping of the waves, the growth rate is calculated from the equation below:

$$\gamma = -Im(\mathcal{D})/(\partial(Re(\mathcal{D}))/\partial\omega) \qquad (11)$$

where $\mathcal{D}$ refers to the dispersion relation in Eq. (3), $Im(\mathcal{D})$ is the imaginary part of $\mathcal{D}$, and $Re(\mathcal{D})$ is the real part of $\mathcal{D}$.

From Eqs. (3) to (11), we can isolate different resonance harmonic numbers $n$ and calculate contributions to the growth rate from isolated resonances. The results are shown in Figures 1(c)-1(e). Both Landau and cyclotron resonances contribute to the unstable region of beam-driven ECH waves in wave number space. The major contributors to wave growth are cyclotron resonance at $n = -1$ and cyclotron resonances at higher resonance harmonic numbers.

Figure 2 is a geometric interpretation of wave-particle interaction in the velocity space of the electron distribution function. Considering an isolated resonance with resonance harmonic number $n$, when analyzing the particle equation of motion for the non-relativistic case using a Hamiltonian approach, we obtain that quantity $C_n$ in Eq. (12) is a constant [Shklyar and Matsumoto, 2009]

$$C_n = nW - \mu\omega = \text{constant} \qquad (12)$$

where $W$ is the particle kinetic energy ($W = \frac{1}{2}m(v_\perp^2 + v_\parallel^2)$), $\mu$ is ($\frac{1}{2}mv_\perp^2)/\omega_{ce}$, and $\omega$ is the wave frequency. In terms of Landau resonance, when $n = 0$, Eq. (12) reduces to $v_\perp = \text{constant}$. Taking into account cyclotron resonance when $n = 1$ and the resonance condition described by Eq. (2), we obtain the "diffusion curve" or "resonance curve" discussed in many papers [Gendrin, 1968, 1981; Summers et al., 1998; Thorne et al., 2005]. When electrons interact with waves at the resonance velocity, they diffuse along the curve corresponding to constant $C_n$, and the net transport in phase space is towards regions with lower phase space density. If this direction is towards smaller particle kinetic energy, particles will lose energy when interacting with waves, resulting in wave generation. When calculating the curve corresponding to the constant $C_n$ in Figure 2 and the resonance velocity with $n = 0$, -1, -2, we used wave parameters corresponding to the maximum growth rate in Figure 1. The plasma parameters for the electron distribution function in Figure 2 are the same as in Figure 1, as shown in Table 1. Taking into account Landau resonance when $n = 0$ and cyclotron resonances when $n = -1$ and $n = -2$, Figure 2 depicts the curves corresponding to constant phase space density, constant $C_n$, and constant energy, where magenta lines with arrows indicate the directions in which particles diffuse. The stability of the wave can only be determined after the combined effects from all perpendicular velocities have been accounted for, which is accomplished by integration along the vertical line for a fixed resonant (parallel) velocity. However, it is evident from Figure 2 that when $n = 0$ (Landau resonance) and when n = -1 and -2 (cyclotron resonances) electrons will lose energy and waves will gain energy due to the overwhelming contribution from the normalized perpendicular velocities of 0.2-0.5, where the downward gradient in phase space density points towards lower energies.

Next, we investigate the excitation of an ECH wave by low-energy electron beams at its second harmonic frequency band (between $2f_{ce}$ and $3f_{ce}$). When the electron distribution functions in Table 1 are used, ECH waves at the second harmonic band are stable. Therefore we increased the drift velocity of electron beam components and used the electron distribution functions listed in Table 2 to solve the dispersion relation for the second harmonic. The background magnetic field strength is 50nT. Figure 3 shows the wave frequency and growth rate in wave number space for the second harmonic band; we can see that growth (within the black contour) can occur over a significant portion of the frequency-wave number space. The most unstable solution at the second harmonic frequency band for the choice of parameters in Table 2 is at wave frequency of $2.12f_{ce}$ and at wave normal angle of 76**º**. We conclude that ECH waves at higher harmonic frequency bands can also be driven unstable by electron beams.

## 3. Parametric study of beam-driven ECH waves

Before presenting the results from our parametric study of beam-driven ECH waves, we would like to discuss electron acoustic waves which can also be driven unstable by cold beams in a warm plasma. During our search for ECH waves with maximum growth rate, we sometimes found that the most unstable wave is in the propagation direction parallel to the magnetic field (see the case in supporting information). We confirmed these to be electron acoustic waves driven unstable through Landau resonance with electron beams [Gary & Tokar, 1985; Singh and Lakhina, 2001; Lu et al., 2005]. Electron acoustic waves often overlap with ECH waves in frequency, making it difficult for us to distinguish between them from their dispersion relations alone. By changing the density and temperature of the cold electron population, we can vary the frequency of an electron acoustic wave and avoid an overlap of the two waves in real frequency. In this parametric survey, electron acoustic waves are not excited or do not appear in the $f_{ce}$ to $2f_{ce}$ frequency range. A wider parameter regime, allowing for both waves to be excited simultaneously needs to be considered in the future.

In this section, we conduct a parametric survey of beam-driven ECH waves using the plasma components listed in Table 1: a hot electron component, a cold electron component, and two electron-beam components streaming in opposite directions. To better understand the dependence of the beam-driven ECH wave growth rate on various plasma parameters, we plot the electron distribution functions as a function of parallel velocity with zero perpendicular velocity for different plasma parameters in Figure 4. The total electron distribution function is defined as

$$f(v_\perp, v_\parallel) = \frac{n_H}{\pi^{2/3}\alpha_{\perp H}^2 \alpha_{\parallel H}} \exp\left(-\frac{v_\parallel^2}{\alpha_{\parallel H}^2}\right) \cdot \exp\left(-\frac{v_\perp^2}{\alpha_{\perp H}^2}\right)$$

$$+ \frac{n_c}{\pi^{2/3}\alpha_{\perp c}^2 \alpha_{\parallel c}} \exp\left(-\frac{v_\parallel^2}{\alpha_{\parallel c}^2}\right) \cdot \exp\left(-\frac{v_\perp^2}{\alpha_{\perp c}^2}\right)$$

$$+ \sum \frac{n_b}{\pi^{2/3}\alpha_{\perp b}^2 \alpha_{\parallel b}} \exp\left(-\frac{(v_\parallel - v_b)^2}{\alpha_{\parallel b}^2}\right) \cdot \exp\left(-\frac{v_\perp^2}{\alpha_{\perp b}^2}\right) \tag{13}$$

where the subscript $H$ is the hot electron population, subscript $c$ is the cold electron population, and subscript $b$ is the electron beam population. We will use this figure during the section to explain the evolution of the phase space gradient magnitude at the various resonant frequencies, as the parameters of the distribution function are modified.

Each figure in this section (Figures 5 through 8) shows the dependence of wave properties on two different sets of plasma parameters. The plasma parameters represented by the horizontal axis and the vertical axis consist of 50 different values, respectively, and there are 50x50 grid points in each figure. For every grid point, we solve the hot plasma dispersion relation for the first harmonic band of beam-driven ECH waves in wave number space, with wave vector $k\rho_e$ ranging from 0 to 30 and wave normal angle ranging from $0^0$ to $90^0$. We search for an ECH wave with maximum growth rate in the wave number space, and every grid point in these figures represents a solution corresponding to the most unstable ECH wave in the first harmonic frequency band.

## 3.1 Dependence on beam density and beam velocity

Figure 5 illustrates the dependence of the wave growth rate, wave frequency, and wave normal angle of the most unstable wave on the beam density and beam velocity (beam density refers to the density of one electron beam component rather than the total density of two electron beam components). We vary the beam density from 0 to $0.1 cm^{-3}$ and the normalized beam velocity (normalized to the thermal velocity of the beam) from 0 to 5 keeping all the other plasma parameters in Table 1 unchanged. Only grids with maximum growth rate greater than 0 (unstable wave) are plotted. At fixed beam velocity, the ECH wave growth rate increases with beam density; at fixed beam density, the ECH wave growth rate increases with beam velocity. This is because when beam density or beam velocity decrease, the electron distribution function "flattens" with subtler beam characteristics and gentler gradients (see Figure 4(b) and Figure 4(c) for illustrations). Therefore, ECH waves stabilize. The wave frequency decreases slightly with beam density. The most unstable ECH waves become more oblique when beam velocity increases and less oblique when beam density increases.

## 3.2 Dependence on beam temperature and beam temperature anisotropy

Next, in Figure 6, the parallel beam temperature is varied from 10eV to 1keV, and the beam temperature anisotropy (defined as the ratio of perpendicular beam temperature to parallel beam temperature) from 0.15 to 0.9 (we change the temperature anisotropy of the beam by varying its perpendicular temperature while keeping its parallel temperature unchanged). All the other plasma parameters remain the same as in Table 1. At fixed beam temperature anisotropy, as the beam temperature increases from 10eV up to ~30eV, the growth rate of ECH wave increases because the waves resonate with a steeper part of the phase space density gradient. When the beam temperature further increases from 30eV to 1keV, the growth rate contributed from cyclotron resonances decreases. This is because the distribution function broadens (see Figure 4(d) for illustrations) and ECH waves stabilize. At fixed beam temperature in the parallel direction, the growth rate of ECH waves decreases when electron beams become more perpendicularly anisotropic. For the most unstable wave, the wave normal angle of ECH waves increases with beam temperature and decreases with beam temperature anisotropy.

## 3.3 Dependence on temperature and density of hot electrons

We vary the temperature of the hot electron population (the first component in Table 1) from about 60eV to 16keV as shown in the horizontal axis in Figure 7. The density of the hot electron population on the vertical axis ranges from $0.05 cm^{-3}$ to $5 cm^{-3}$; the ratio of the electron plasma frequency to the electron cyclotron frequency $\omega_{pe}/\omega_{ce}$ ranges from 2.6 to 14.5. The electron plasma beta in this parametric space ranges from 0.002 to 13. At fixed hot electron temperature, the growth rate of the ECH wave decreases as the hot electron density increases. This is because the electron distribution function flattens with a smaller phase space density gradient in the parallel direction when the density of hot electrons increases (see Figure 4(f) for illustration). The growth rate contributed from cyclotron resonances decreases, and ECH waves are thus damped by Landau resonance with the hot electron component. When the hot electron temperature decreases at fixed hot electron density, the electron distribution function broadens and the growth rate of ECH waves decreases (see Figure 4(g) for illustration). Towards the regions of lower growth rate with larger hot electron density and lower hot electron temperature in the parametric space, ECH wave normal angles for the most unstable waves become larger.

## 3.4 Dependence on temperature and density of cold electron

Next we vary the density and temperature of the cold electrons (the second component in Table 1) leaving the other plasma parameters the same as in Table 1 (beam density is changed to $0.025 cm^{-3}$). We vary the density of the cold electron component from $0.005 cm^{-3}$ to $0.5 cm^{-3}$, and the temperature of the cold electron component from 0.1eV to 10eV. When the temperature of the cold electron component becomes too high, ECH waves stabilize because ECH waves are Landau damped due to their interaction with cold electrons (as illustrated in Figure 4(h)). When the density of the cold electron component is too large ($> 0.1 \ cm^{-3}$) (or too small ($< 0.01 \ cm^{-3}$)), the frequency of a beam-driven ECH wave gets close to $f_{ce}$ (or to $2f_{ce}$) and the resonance velocity in Eq. (2) for $n = +1$ ($n = +2$) becomes too small. Under those conditions, ECH waves are damped by cyclotron resonance with the $n = +1$ (or n = +2) resonance. Notably, we find that the wave frequency of ECH waves at their first harmonic changes from $1.1f_{ce}$ to $1.9f_{ce}$ when the density of the cold electron component changes. When we vary the plasma parameters for electron beams and hot electrons, however, the frequency of ECH waves remains nearly unchanged. Therefore, electron density is a very important parameter in determining the frequency of beam-driven ECH waves. Conversely, wave frequency measurements are an important diagnostic of cold electron density, which is often poorly constrained.

# 4. Summary and Discussion

Using an electron distribution function with a hot component, a cold component, and two beam components as input parameters, we solved the hot plasma dispersion relation for ECH waves and investigated wave generation and wave properties under different plasma conditions. Our primary findings are summarized below:

1. At moderately oblique wave normal angles, ECH waves driven by low-energy electron beams are unstable. The growth rate of beam-driven ECH waves is mainly controlled by cyclotron resonance when the resonance harmonic number $n$ is -1 and by cyclotron resonances when $n$ is between -2 and -5.
2. Electron cyclotron harmonic waves at their second harmonic frequency band can also be driven unstable by electron beams.
3. The maximum growth rate of a beam-driven ECH wave increases with electron beam density and electron beam drift velocity. The wave normal angle of a beam-driven ECH wave increases with beam velocity and decreases with beam density.
4. The maximum growth rate of a beam-driven ECH wave decreases with beam temperature and beam temperature anisotropy. The wave normal angle increases with beam temperature and decreases with beam temperature anisotropy.
5. When the hot electron density is higher, the growth rate of a beam-driven ECH wave is lower and its wave normal angle is larger. When the hot electron temperature is higher, the growth rate of a beam-driven ECH wave is higher and its wave normal angle is smaller.

6. A beam-driven ECH wave is stabilized when the temperature of the cold electron component is too large and when the density of the cold electron component is either too large or too small. The cold electron density controls the frequency of the most unstable beam-driven ECH waves.

Our results reveal the nature of the beam-driven ECH wave excitation mechanism and demonstrate the dependence of wave properties on various plasma parameters. The loss-cone distribution is not the only free energy source for ECH waves: they can also be driven unstable by electron beams for a wide range of ambient plasma parameters that encompass the magnetotail plasma sheet. Such an excitation mechanism has been confirmed observationally in Zhang et al [2021] and further explored theoretically here. Therefore, our work improves our understanding of this previously unknown excitation mechanism.

From our parametric survey, we found that electron acoustic waves can also be excited by low-energy electron beams in a similar frequency range as ECH waves, but they propagate mostly parallel to the magnetic field (see supporting information). Excited by beam-plasma instability, electron acoustic waves may coexist with beam-driven ECH waves in observations [Roeder et al., 1991] and may also compete with beam-driven ECH waves by relaxing electron beams upon saturation [Omura and Matsumoto, 1987; An et al., 2017 ; Agapitov et al., 2018].

Beam-driven ECH waves with wave frequency ranging from $1.1f_{ce}$ to $1.9f_{ce}$ and wave normal angles ranging from $40^0$ to $80^0$ are unstable under a wide range of plasma conditions in electron plasma betas from as small as 0.003 and as large as 12.9 (as shown in Figure 7). Excitation of beam-driven ECH waves under such a wide range of plasma parameters suggests that they might exist not only in the magnetotail but also in many other regions of Earth's magnetosphere. Likely to originate in the ionosphere, low-energy electron beams could be (1) secondary electrons produced by precipitating plasma sheet electrons [Khazanov et al., 2014; Artemyev et al., 2020]; (2) upward electron beams accelerated by an electric field parallel to the magnetic field and near downward field-aligned currents [Carlson et al., 1998; Hull et al. 2020]. Such low-energy ionospheric electron beams have been observed in both the magnetotail [Walsh et al., 2013; Artemyev et al., 2015] and the outer radiation belt [Kellogg et al., 2011; Mourenas et al., 2015]. Therefore, we would expect beam-driven ECH waves to exist not only near dipolarization fronts in the plasma sheet, as shown by Zhang et al. [2021], but also in the magnetotail and in the inner magnetosphere, even during quiet times.

Electron cyclotron harmonic waves driven unstable by a loss-cone distribution resonate with electrons through cyclotron resonance when the resonance harmonic number, $n$, is 1; beam-driven ECH waves are driven unstable through cyclotron resonance with electron beams when $n$ is -1 and of higher order. Excited by a totally different mechanism from the one that excites loss cone-driven ECH waves, beam-driven ECH waves would interact with electrons in a completely different way. During the excitation of beam-driven ECH waves, energy is transferred from electron beams to ECH waves and ECH waves saturate by slowing down electron beams. The electron distribution function would eventually flatten and form a plateau in the velocity space. Additionally, beam-driven ECH waves are characterized by a moderately oblique wave normal angle. Because wave normal angle is an important parameter in evaluating the pitch-angle diffusion coefficients for ECH waves, the pitch-angle diffusion coefficient profile as a function of equatorial pitch angle and energy for beam-driven ECH waves would be different from the profile for loss cone-driven ECH waves. Evaluating the effects of beam-driven ECH waves on electron dynamics is thus important to explore in future studies.

Acknowledgements

This work was supported by NASA contract NAS5-02099. We thank J. Hohl for help with editing. Anton Artemyev and Xiao-Jia Zhang are also supported by NSF GEM grants #1902699 and #2026375. Xiao-Jia Zhang acknowledges NASA grant 80NSSC18K1112.

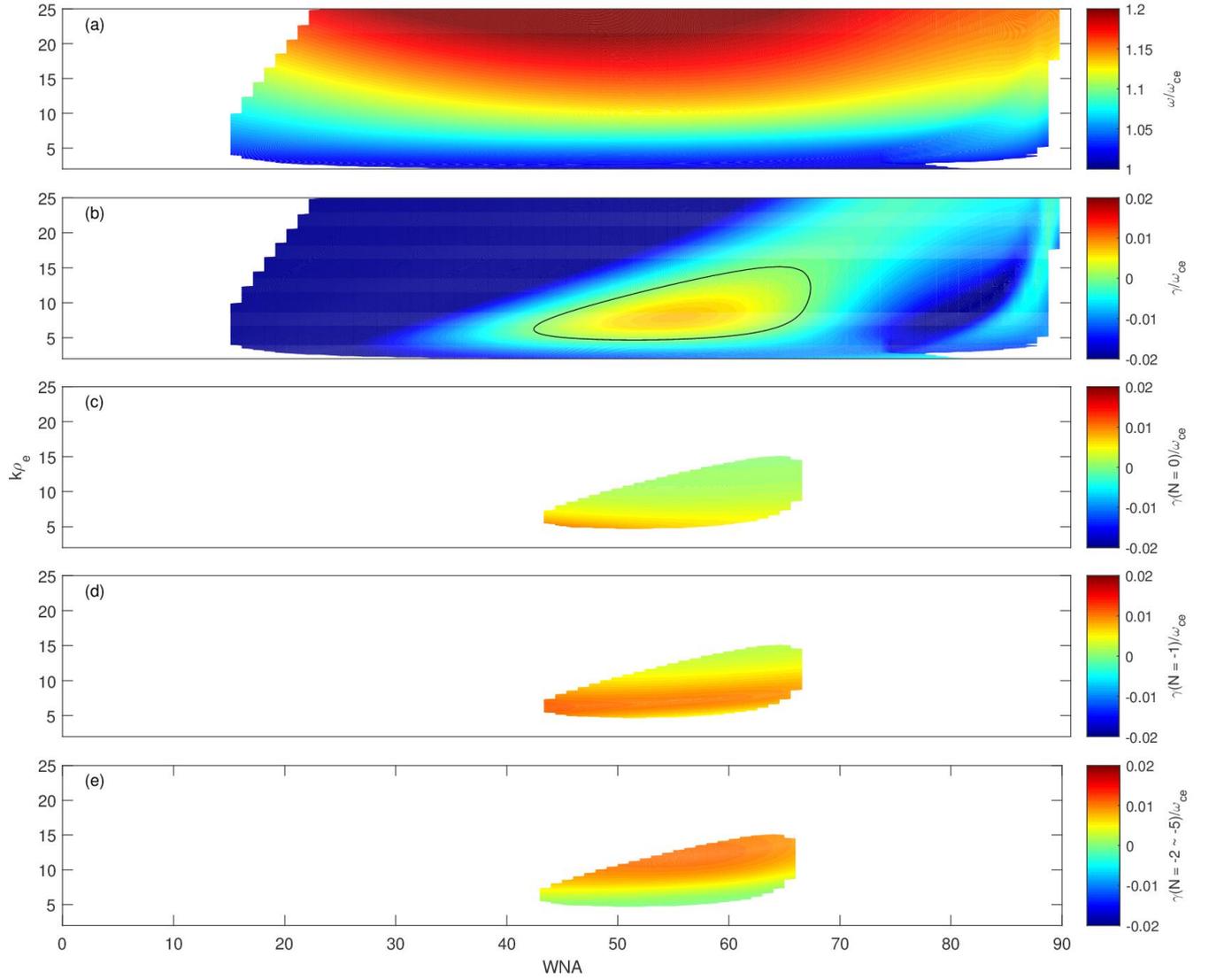

**Figure 1**: Dispersion relations and contributions to the growth rate from different resonance harmonic numbers for beam-driven ECH waves. The horizontal axis is the wave normal angle, and the vertical axis is the wave vector normalized to the gyroradius of the hot electron component (the first component in Table 1). (a): Wave frequency normalized to the electron cyclotron frequency; (b): Growth rate normalized to the electron cyclotron frequency (the black line indicates the zero growth rate contour); (c): Growth rate contributed by the Landau resonance when $n = 0$. Only data points with positive growth rate are plotted; (d): Growth rate contributed by the cyclotron resonance when $n = -1$; (e): Growth rate contributed by the cyclotron resonances at higher orders (summation of $n = -2,-3,-4,-5$)

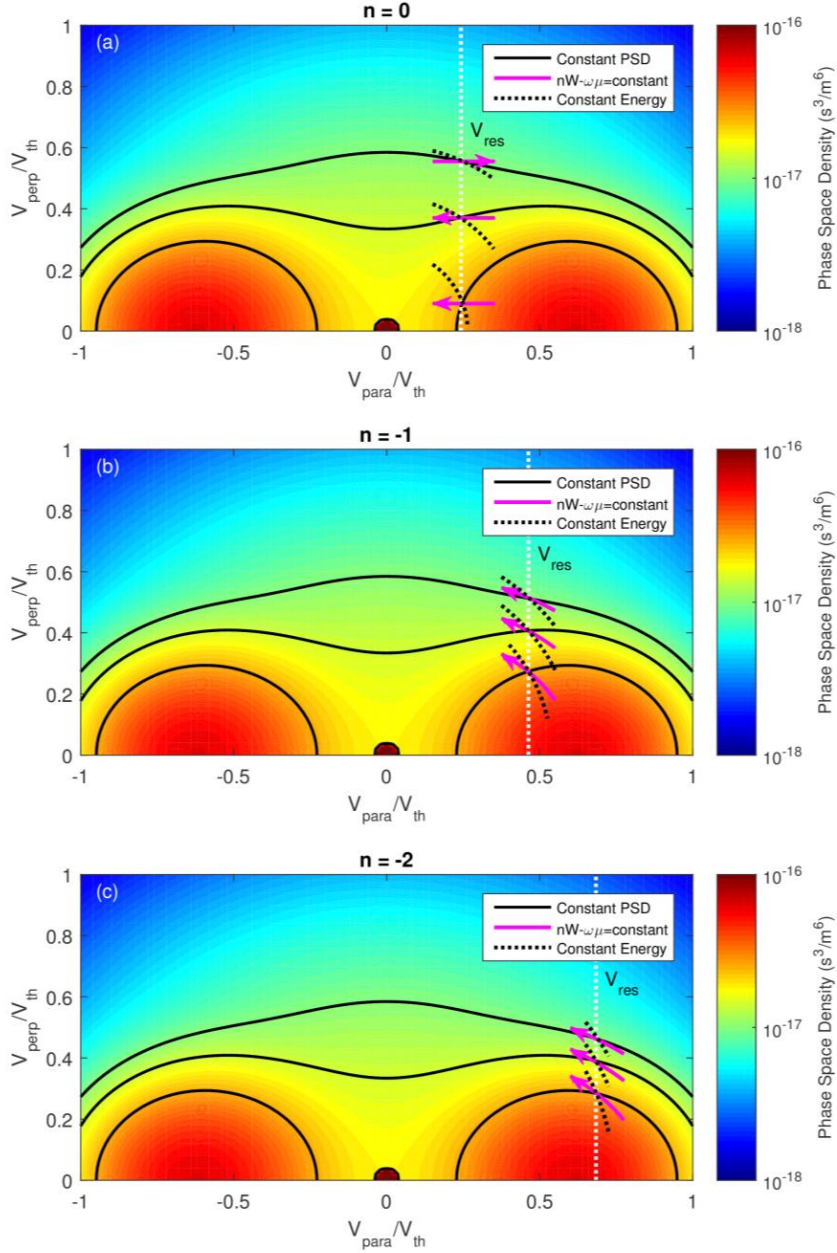

**Figure 2**: The electron distribution function plotted using the plasma parameters listed in Table 1. Parallel velocity on the horizontal axis and perpendicular velocity on the vertical axis are normalized to the thermal velocity of the first component in Table 1. The solid black lines represent constant phase space density, and the dotted black lines represent constant particle energy. Using the wave properties corresponding to the maximum growth rate in Figure 1, we plot the contours of Eq. (12), indicated by solid magenta lines with arrows, and the resonance velocity, indicated by dotted white lines. Figures 2(a), 2(b), and 2(c) demonstrate the case when $n$ = 0, -1, and -2 respectively.

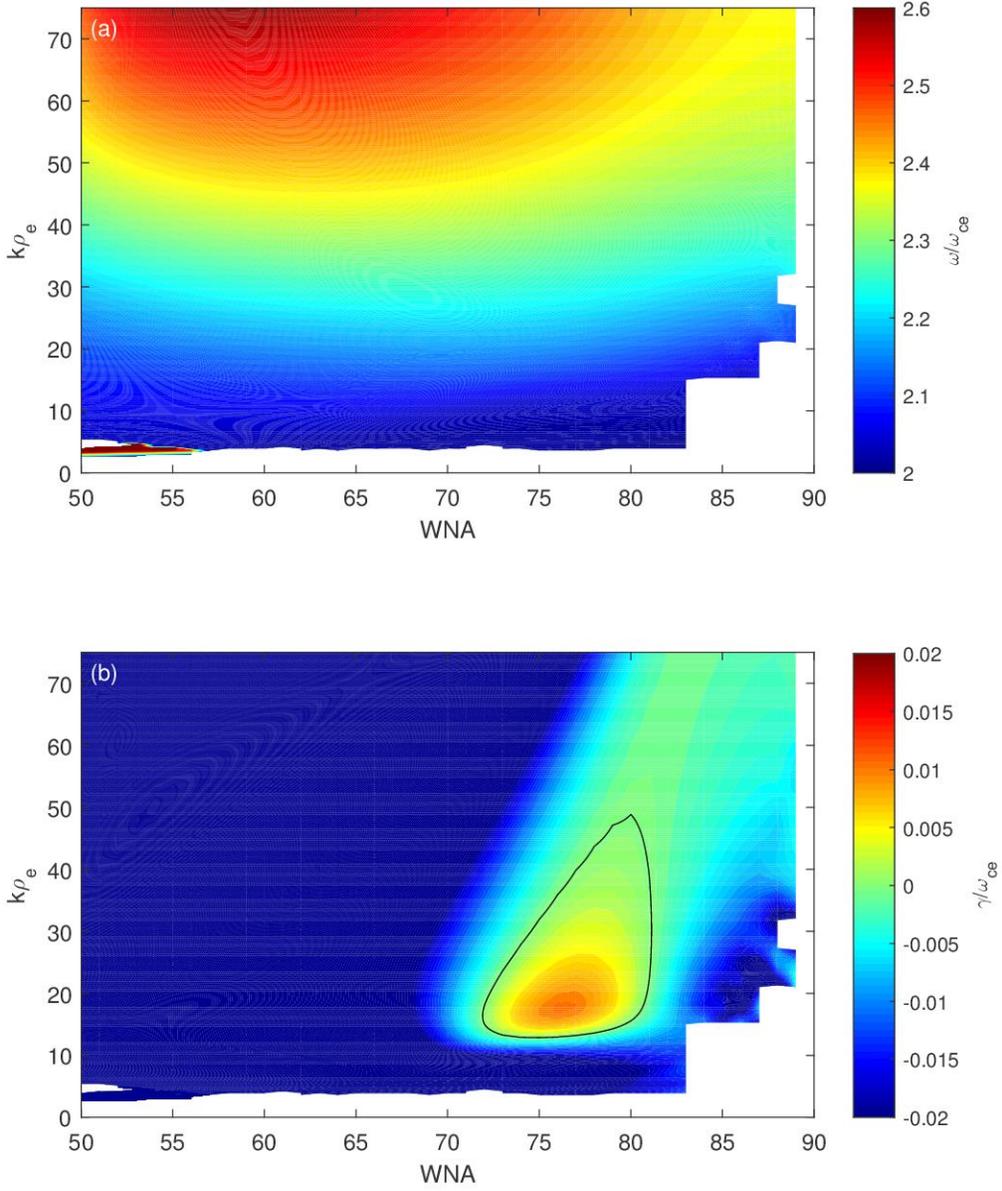

**Figure 3**: Dispersion relations and the growth rate for beam-driven ECH waves at the second harmonic frequency band. Input plasma parameters are listed in Table 2. The horizontal axis is the wave normal angle; the vertical axis is the wave vector normalized to the gyroradius of the first component in Table 2. Figures 3(a) and 3(b) show the wave frequency and growth rate normalized to the electron cyclotron frequency.

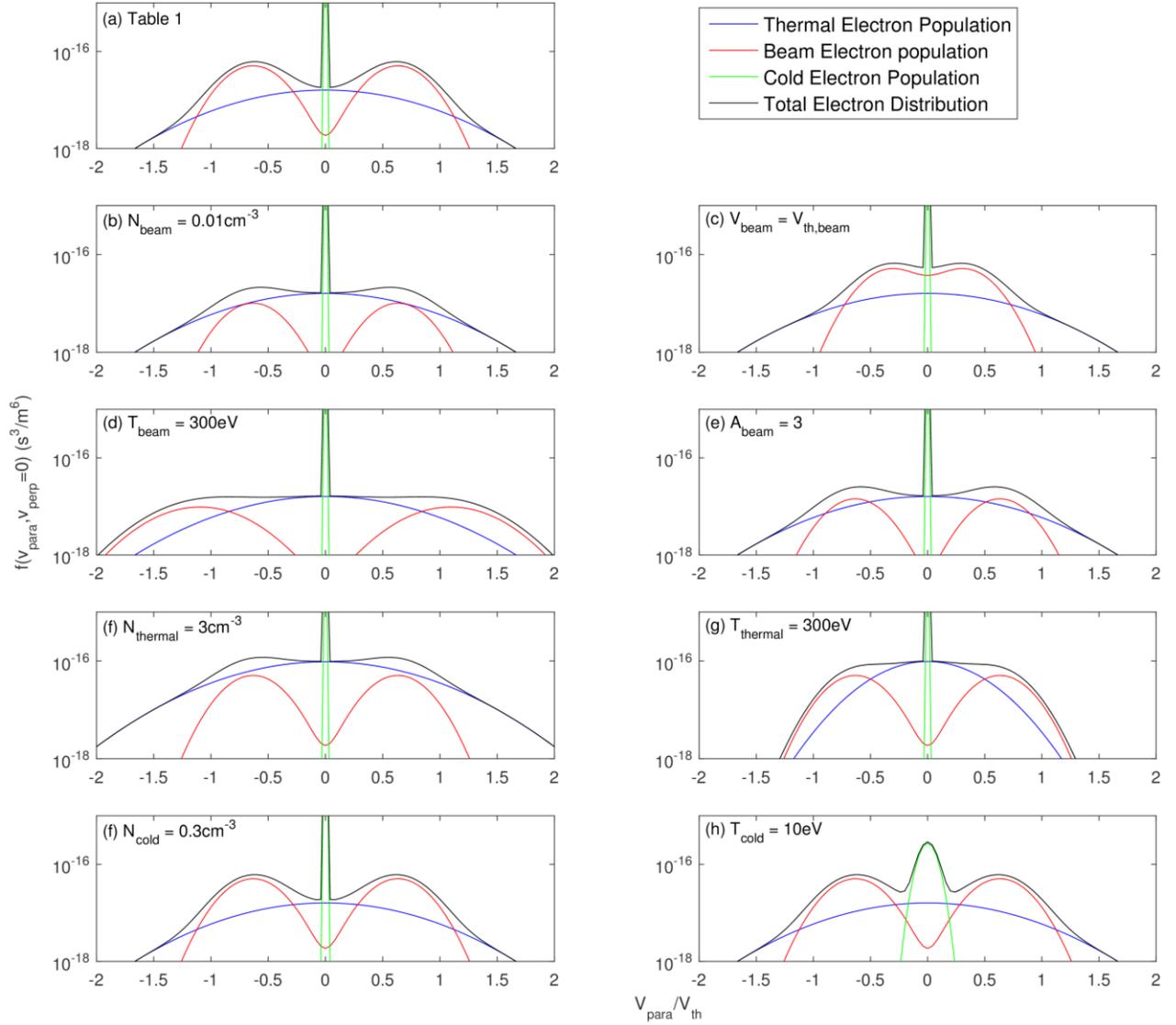

**Figure 4**: The electron phase space density as a function of parallel velocity with zero perpendicular velocity. The parallel velocity is normalized to the thermal velocity of a 1keV electron. We use the electron plasma parameters listed in Table 1 and vary a parameter at one time for each figure. (a): Plasma parameters in Table 1; (b): The beam density is changed to $0.01cm^{-3}$; (c): The normalized beam velocity is changed to 1; (d): The beam temperature is changed to 300eV; (e): The temperature anisotropy is changed to 3; (f): The hot electron density is changed to $3cm^{-3}$; (g): The hot electron temperature is changed to 300eV; (f): The cold electron density is changed to $0.3cm^{-3}$; (h): The cold electron temperature is changed to 10eV.

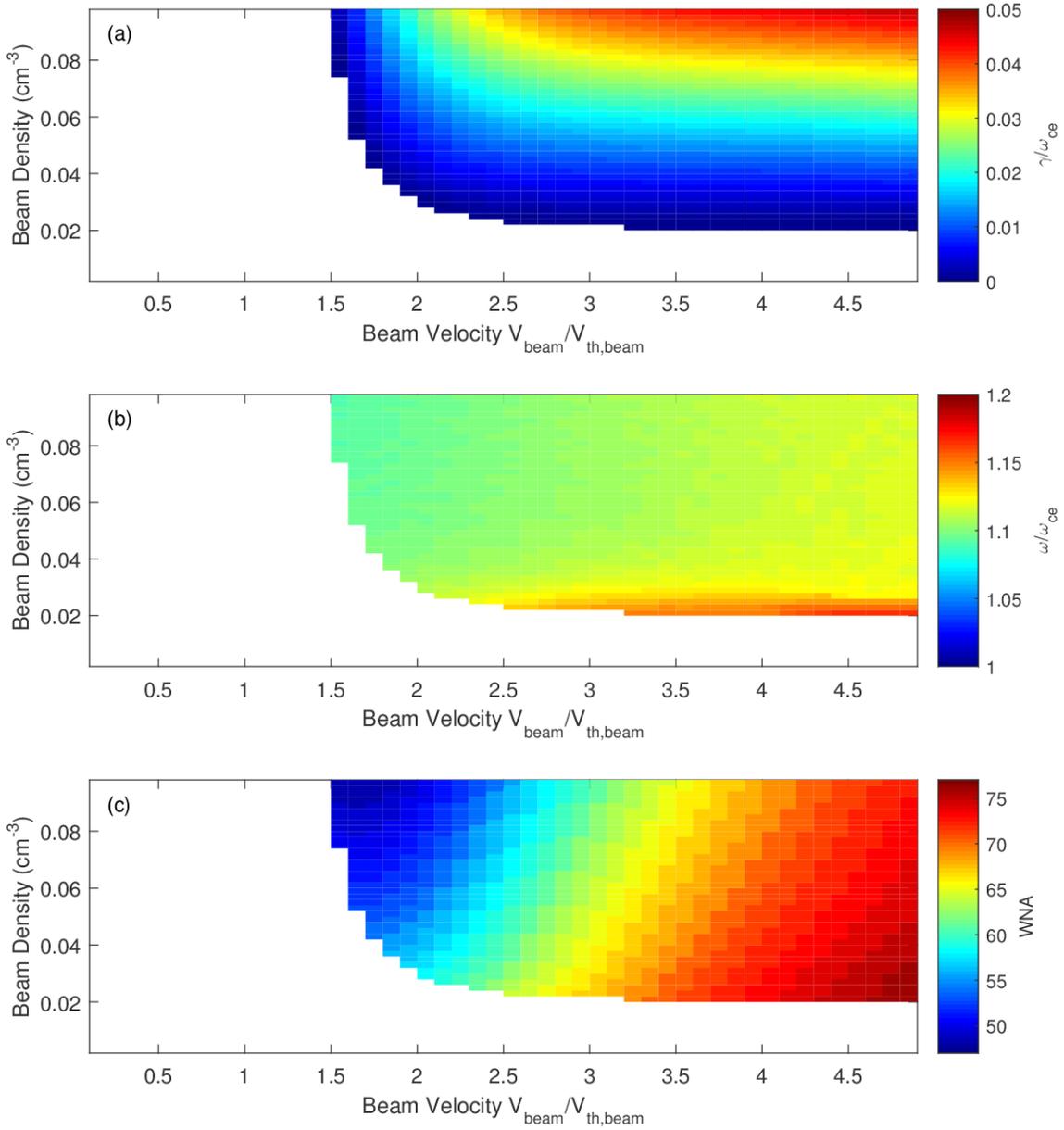

**Figure 5**: A total of 50x50 data points are in this parameter space. Each data point represents wave properties corresponding to the maximum growth rate. (a): Normalized growth rate as a function of beam velocity and beam density. The beam velocity is normalized to the thermal velocity of the beam; (b): The wave frequency is normalized to the electron cyclotron frequency; (c): Wave normal angle.

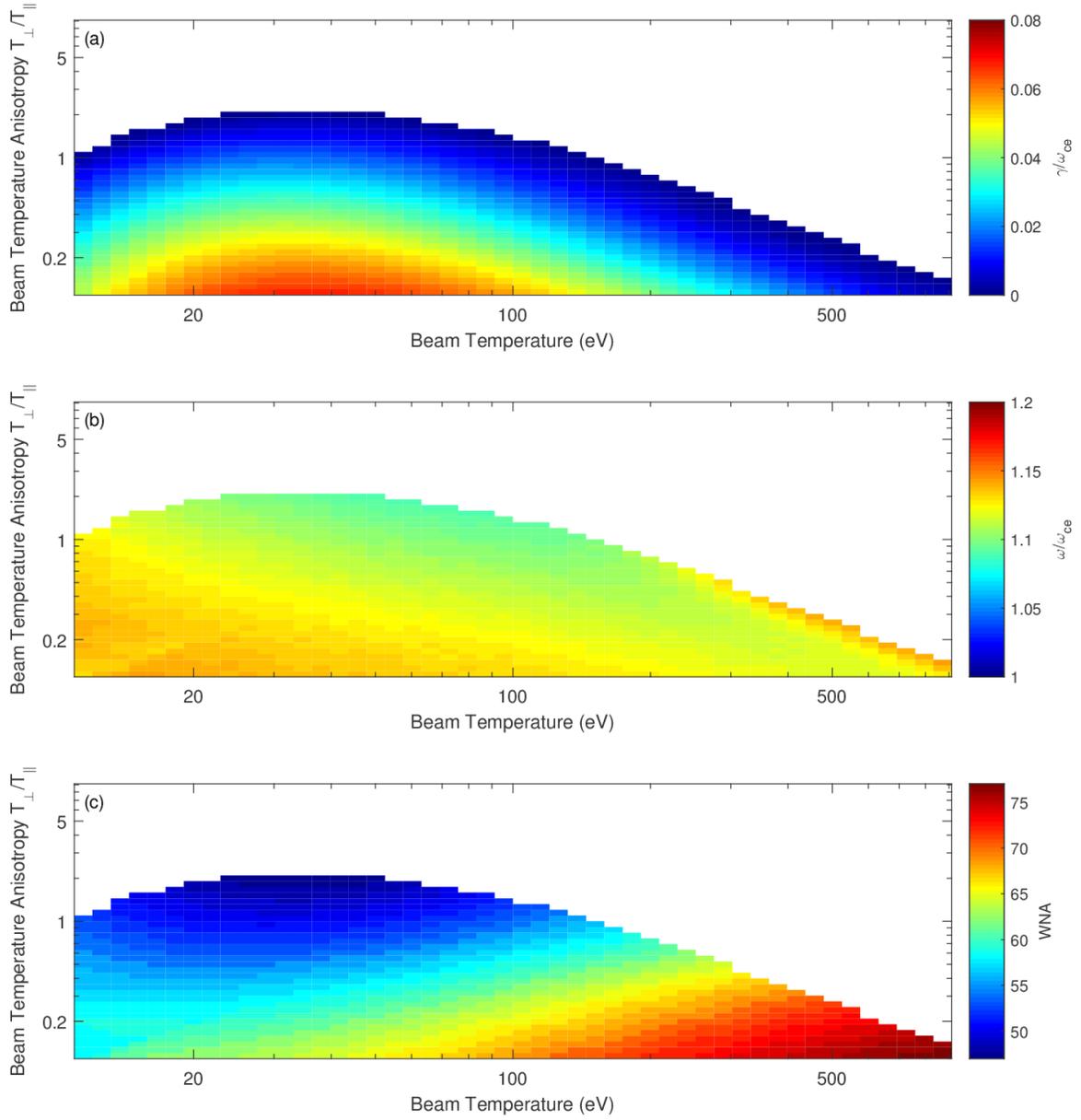

**Figure 6**: The figure format is the same as Figure 4. Horizontal axis is beam temperature in the parallel direction and vertical axis is beam temperature anisotropy defined as the ratio between perpendicular and parallel beam temperature.

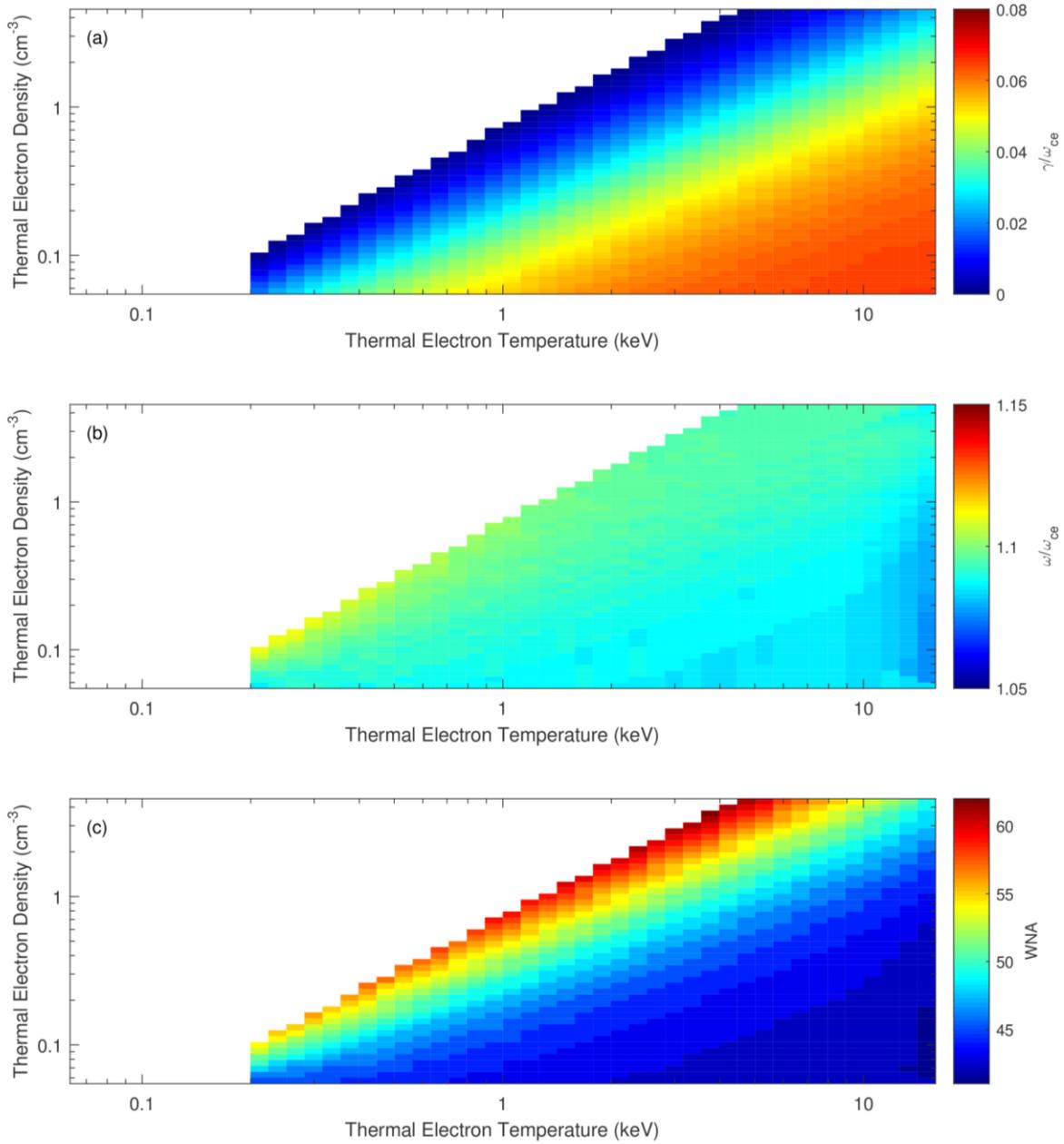

**Figure 7**: The figure format is the same as in Figure 4. The horizontal axis is the temperature of the hot electron component; the vertical axis is the density of the hot electron component.

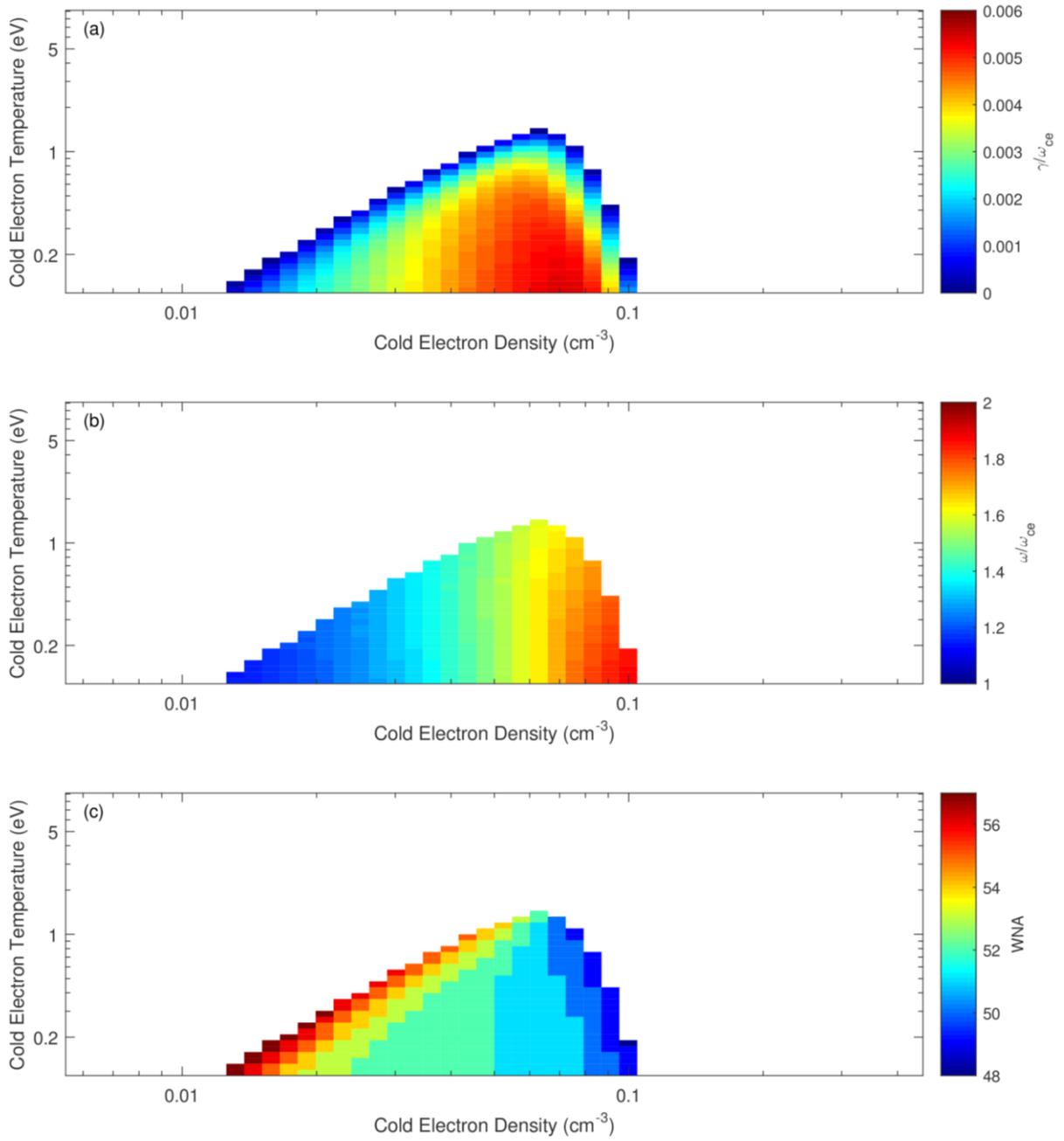

**Figure 8**: The figure format is the same as in Figure 4. The horizontal axis is the density of the cold electron component, and the vertical axis is the temperature of the cold electron component.

Table 1 Electron distribution function

| Component | n ($cm^{-3}$) | $T_{\parallel}$ (eV) | $T_{\perp}/T_{\parallel}$ | $V_{drift}/V_{thermal}$ |
|---|---|---|---|---|
| 1 Hot electron | 0.5 | 1000 | 0.85 | 0 |
| 2 Cold electron | 0.01 | 0.1 | 1 | 0 |
| 3 Electron beam | 0.05 | 100 | 0.85 | 2 |
| 4 Electron beam | 0.05 | 100 | 0.85 | -2 |

Table 2 Electron distribution function

| Component | n ($cm^{-3}$) | $T_{\parallel}$ (eV) | $T_{\perp}/T_{\parallel}$ | $V_{drift}/V_{thermal}$ |
|---|---|---|---|---|
| 1 Hot electron | 0.5 | 1000 | 0.85 | 0 |
| 2 Cold electron | 0.08 | 0.2 | 1 | 0 |
| 3 Electron beam | 0.05 | 100 | 0.85 | 5 |
| 4 Electron beam | 0.05 | 100 | 0.85 | -5 |